\documentclass[preprint,12pt]{elsarticle}
\usepackage{graphicx,amsmath,bbm,bm,array,subfigure}
\usepackage{mathtools}
\usepackage{float}
\usepackage[linktocpage=true,colorlinks=true,linkcolor=blue,citecolor=blue]{hyperref}
\journal{Nuclear Physics A}
\begin{document}
    
\begin{frontmatter}	
	\title {Hadron resonance gas model with repulsive mean-field interactions:  specific heat, isothermal compressibility and speed of sound}
	\author{Somenath Pal\footnote{somenathpal1@gmail.com}}
	\address{Department of Physics, University of Calcutta, 92, A.P.C. Road, 
	Kolkata-700009, India}
     
    \author{Guruprasad Kadam\footnote{guruprasadkadam18@gmail.com}}
    \address{School of Physical Sciences,  National Institute of Science Education and Research Bhubaneswar,  HBNI,  Jatni 752050, Odisha, India}
	
	\author{Abhijit Bhattacharyya\footnote{abhattacharyyacu@gmail.com}}
	\address{Department of Physics, University of Calcutta, 92, A.P.C. Road, Kolkata-700009, India}
	
	\def\be{\begin{equation}}
	\def\ee{\end{equation}}
	\def\bearr{\begin{eqnarray}}
	\def\eearr{\end{eqnarray}}
	\def\zbf#1{{\bf {#1}}}
	\def\bfm#1{\mbox{\boldmath $#1$}}
	\def\hf{\frac{1}{2}}
	\def\sl{\hspace{-0.15cm}/}
	\def\omit#1{_{\!\rlap{$\scriptscriptstyle \backslash$}
			{\scriptscriptstyle #1}}}
	\def\vec#1{\mathchoice
		{\mbox{\boldmath $#1$}}
		{\mbox{\boldmath $#1$}}
		{\mbox{\boldmath $\scriptstyle #1$}}
		{\mbox{\boldmath $\scriptscriptstyle #1$}}
	}

	\begin{abstract}
		We investigate the effect of repulsive interaction between hadrons on the specific heat ($C_V$), isothermal compressibility ($\kappa_{T}$) and the speed of sound ($C_s^2$) of hot and dense hadronic matter.  
		The repulsive interactions are included through  a mean-field approach where the single particle energy picks correction term  due to mean field interactions between hadrons. This correction term is proportional to the number density of hadrons.  We assume different mean-field interactions for mesons and baryons. We also confront $C_V$ and $C_s^2$ with existing lattice QCD simulation results. We find that the repulsive interactions have very strong effect on $C_V$ and $C_s^2$ while its effect on $\kappa_T$ is very mild.  
	    We finally discuss the implications of our results in the context of heavy-ion collision experiments.  
	\end{abstract}
	

\end{frontmatter}

	\section{Introduction}
	Properties of strongly interacting matter under extreme conditions of high temperature and/or high density is a hot topic of research for many years. Quantum Chromodynamics (QCD), the theory of strong interaction,  predicts that there should be a phase transition from hadronic phase to quark-gluon plasma (QGP) phase at high temperature and/or baryon chemical potential. In spite of many rigorous efforts, dedicated to the study of such system for many years, the phase structure of strongly interacting matter remains somewhat unclear. While the lattice QCD results suggest that there should be a crossover at negligible baryon chemical potential~\cite{Aoki:2006we,Brown:1990ev,Bazavov:2014pvz},  various effective models respecting important aspects of QCD predict the phase transition to be of first order~\cite{Gottlieb:1985ug,Fukugita:1989yb,Schaefer:2004en,Herpay:2005yr,Pisarski:1983ms,Halasz:1998qr,Hatta:2002sj,Bowman:2008kc,PNJL1,PNJL2,PNJL3,PNJL4,PNJL4a,PQM1,PQM2,PQM3,PQM4} at high baryon chemical potential. This first order phase transition should end somewhere in the phase diagram and hence identifying the location of the Critical End Point (CEP) in the phase diagram has been an active field of research~\cite{Fodor:2004nz,Gavai:2004sd,PNJL5,Schaefer:2006ds}. On the other hand, statistical hadronic models, especially hadron resonance gas (HRG) model~\cite{HRG1,BraunMunzinger:1994xr,Cleymans:1996cd,BraunMunzinger:1999qy,Cleymans:1999st,Becattini:2005xt,Andronic:2008gu}, provide quite a satisfactory description of strongly interacting matter at high temperature and moderate density. Ongoing experimental programs at facilities like Large Hadron Collider (LHC) at CERN and Relativistic Heavy-Ion Collider (RHIC) at BNL have revealed many aspects of QCD physics and will continue to do so in future. Future facilities at GSI, Darmstadt and JINR, Dubna will shed more light on the phase diagram of QCD. 
	
	Hadron resonance gas (HRG) model~\cite{HRG1,BraunMunzinger:1994xr,Cleymans:1996cd,BraunMunzinger:1999qy,Cleymans:1999st,Becattini:2005xt,Andronic:2008gu} is quite successful in reproducing particle multiplicity ratios in heavy-ion collision experiments~\cite{BraunMunzinger:1994xr,Andronic:2005yp,Cleymans:2000ck,BraunMunzinger:2001ip,Cleymans:1997sw,Cleymans:1998yb,Becattini:1997ii}. This model is based on Dashen, Ma and Bernstein theorem~\cite{Dashen:1969ep} which states that the attractive interactions in a dilute hadron gas can be mimicked by considering the higher resonance particles as stable particles: this idea is justified in relativistic virial expansion based on S-matrix approach. In addition to attractive interaction among the hadrons, repulsive interactions are also important~\cite{Begun:2012rf}.  Excluded volume hadron resonance gas model (EVHRG) is an extension of HRG model and takes into account short range repulsive interaction among hadrons by considering a non-penetrable excluded volume around the hadrons~\cite{Hagedorn:1980kb,Rischke:1991ke,Cleymans:1992jz,Singh:1991np,Yen:1997rv,Andronic:2012ut,Fu:2012zzc,Bhattacharyya:2013oya}. Another alternative  model which incorporates repulsive interactions among hadrons is HRG mean-field (HRGMF) model which takes into account the repulsive interaction among the hadrons in a system by considering a shift in the single particle energies proportional to the total number density of hadrons~\cite{Kapusta:1982qd,Olive:1980dy}. This relativistic mean-field approach has been used to calculate the
	fluctuations of conserved charges~\cite{Huovinen:2017ogf,Pal:2020ucy} and the transport properties of hot and dense hadronic matter~\cite{Kadam:2019peo}.
	
	Behaviour of certain thermodynamic observables like specific heat ($C_V$), isothermal compressibility ($\kappa_T$) and speed of sound ($C_s^2$) provide useful information about the nature of the phase transition of the system. Specific heat $C_V$ is the change in internal energy for unit change in temperature. This quantity is important in the context of heavy-ion collision phenomenology as it is related to the temperature fluctuations of the system produced in heavy-ion collision by means of event-by-event analysis~\cite{Stodolsky:1995ds,Shuryak:1997yj,Basu:2016ibk}.  $C_V$ is expected to show a jump near a first order phase transition and a singularity for a second order phase transition. Isothermal compressibility represents rate of change of volume of the system with respect to pressure at constant temperature. It is the second order derivative of Gibbs free energy and should diverge at CEP. Fluctuations in a system leads to deviation from equilibrium and, for a brief time, a non-equilibrium condition is reached~\cite{Khuntia:2018non}.  
	Speed of sound reflects the small perturbations produced in the system in its local rest frame. Hence it shows critical behaviour near phase transition i.e. a minimum near phase transition~\cite{Tiwari:2011km}. It is interesting to explore what results are obtained for these quantities in HRGMF model.
	
	In this work our aim is to study the effect of repulsive interactions on the thermodynamic quantities, namely specific heat, isothermal compressibility and speed of sound within the ambit of extended HRG model in which the repulsive interactions are accounted via  mean-field approach.  All these quantities are sensitive to the order of phase transition and hence they are important observables in the context of heavy-ion collision experiments. In the current literature these quantities have been studied either in the case of ideal HRG approximation or separately within  the ambit of  excluded volume HRG model. In this work we provide unified study of all  these quantities within  the ambit of  HRGMF model which includes repulsive interactions and compare our results with recent LQCD simulation results at zero as well as at finite baryon chemical potential. We shall see that the repulsive interaction plays a very important role in describing the thermodynamics of hadronic matter especially near quark-hadron transition temperature $T_c$. We shall then discuss the importance of repulsive interactions in the context of heavy-ion collisions. For this purpose we shall discuss the beam-energy dependence of $C_V, \kappa_T$ and $C_s^2$ which are important thermodynamic observables in HIC experiments. It is important to note that our purpose in this work is not to extract $C_V, \kappa_T$ and $C_s^2$ from the experimental data, but to estimate the magnitude of these quantities within  the ambit of HRGMF model along the freeze-out curve parametrised using statistical thermal models.  
	
	We organise the paper as follows. In Sec.~\ref{secII} we shall recapitulate the HRGMF model. In Sec.~\ref{secIII} we discuss the thermodynamics of HRGMF model with special emphasis on  the results of specific heat, isothermal compressibility and speed of sound at zero as well as at finite baryon chemical potential. Sec.~\ref{secIV} contains a discussion in the context of heavy ion collisions.  In Sec.~\ref{secV} we  summarise and conclude.
	
	\section{ Hadron resonance gas model with repulsive mean-field interactions }
	\label{secII}
	In this section we will recapitulate  HRGMF model. The details of the model can be found in reference~\cite{Kadam:2019peo}. The basic thermodynamic quantity that describes the hadron resonance gas is the partition function 
	\begin{eqnarray}
	\text{ln}\mathcal{Z_{\text{HRG}}}(T,\mu_B, V)= \! \!  \! \! \sum  \limits_{i\in \text{mesons}}  \! \! \! \!\text{ln}\mathcal{Z}_i(T)
	+\sum_{i\in \text{baryons}}
	\text{ln}\mathcal{Z}_i(T,\mu_B)
	\end{eqnarray}
	where $\mu_B$ is  the baryon chemical potential. The partition function of $i^{\text{th}}$ hadronic species is 
	\begin{equation}
	\text{ln}\mathcal{Z}_i(T,V,\mu_B)=\pm V\int d\Gamma_{i}\: \text{ln}[1\pm e^{-\frac{(E_i-B_i\mu_B)}{T}}]
	\end{equation}
	where, for the $\text{i}^{th}$ species of hadrons, $g_i$ is the degeneracy,  $d\Gamma_{a}\equiv\frac{g_{a}d^{3}p}{(2\pi)^3}$, $E_{i}=\sqrt{p^2+m_i^2} \ $ is the single particle energy with mass $m_i$.  Upper (lower) sign corresponds to fermions (bosons). 
	This Ideal HRG model can be extended by including short range repulsive interactions between hadrons via mean-field approach where the single particle energies  get modified by the mean-field repulsive interaction as ~\cite{ Kapusta:1982qd,Olive:1980dy}
	\begin{equation}
	\varepsilon_{i}=\sqrt{p^2+m_{i}^2}+\mathcal{U}(n)=E_{i}+\mathcal{U}(n)
	\label{dispersion}
	\end{equation}
	where  $n$ is the total hadron number density and $\mathcal{U}$ represents the potential energy  between hadrons which  is taken to be proportional to total hadron density $n$, $\mathcal{U}=Kn$. Here, $K$ is a  phenomenological parameter.
	In the present investigation, we  assume repulsive interaction among meson-meson pairs, baryon-baryon pairs and antibaryon-antibaryon pairs only. Thus in the case of mesons, $n=n_M$ which contains contribution from all the mesons and anti-mesons. Similar considerations for baryons and anti-baryons separately are taken into account. We denote the mean field parameter for baryons  and anti-baryons  by $K_B$, while for mesons we denote it by  $K_M$. Number density of baryon is given by

\begin{equation}
n_{{B}}  (T,\mu_B)=\sum_{i\in B}  \int d\Gamma_{i} \:\frac{1}{e^{\frac{(E_{i}-{\mu_{\text{eff},B}})}{T}}+1}
\label{numdenbaryon}
\end{equation}
where $\mu_{\text{eff},B}=B_i\mu_B-K_Bn_{B}$. The sum is over all the baryons. 
Similarly, the number density of antibaryons is
\begin{equation}
n_{{\bar B}}  (T,\mu_B)=\sum_{i\in {\bar B}}  \int d\Gamma_{i} \:\frac{1}{e^{\frac{(E_{i}-{\mu_{\text{eff},\bar B}})}{T}}+1}
\label{numdenbaryon}
\end{equation}
The form of the effective chemical potential for antibaryons is same as that of baryons however the numerical value 
changes as the baryon number for the antibaryons are negative. Note that repulsive mean-field parameter is same 
for baryons as well as anti-baryons.
 For mesons,

\begin{equation}
n_{{M}}(T)=\sum_{i\in M}\int d\Gamma_{i}\:\frac{1}{e^{\frac{(E_{i}-{\mu_{\text{eff},M}})}{T}}-1}
\label{numdenmeson}
\end{equation}
where $\mu_{\text{eff},{M}}=-K_Mn_{M}$ and the sum is over all the mesons. Eqs. (\ref{numdenbaryon})-(\ref{numdenmeson})  are self consistent equations for number density which can be solved numerically.

	The partition function of HRG mean-field  model (HRGMF) is written as,
	\begin{eqnarray}
\text{ln} \ \mathcal{Z_{\text{HRGMF}}}(T,\mu_B, V)=\pm V\sum_{i}\int d\Gamma_{i}\:
	\text{ln}\bigg[1\pm e^{-(\frac{E_i-\mu_\text{eff}}{T})}\bigg]-\phi_{M,B,\bar{B}}(n_{M,B,\bar{B}})
	\end{eqnarray}
where the correction factor $\phi$ is given by
	\begin{equation}
	\phi_{B\{\bar{B}\}}(n_{B\{\bar{B}\}})=-\frac{1}{2}K_Bn_{B\{\bar{B}\}}^2, \hspace{0.5cm}\text{(Baryons)}
	\end{equation}
	and
	\begin{equation}
	\phi_M(n_M)=-\frac{1}{2}K_Mn_M^2,  \hspace{0.5cm}\text{(Mesons)}
	\end{equation}
This correction factor is added in order to avoid the double counting of the potential and it renders the correct energy per particle ($\varepsilon_i=\frac{\partial \epsilon_i}{\partial n_i}$). Thermodynamic quantities can be readily obtained from the partition function. For instance, the expressions for the pressures  for  baryons and mesons are given by

\begin{equation}
P_{B\{\bar B \}} (T,\mu_B)=T\sum_{i\in B \{\bar B\} }\int  d\Gamma_{i} \ \text{ln}\bigg[1+ e^{-(\frac{E_i-\mu_{\text{eff},B\{\bar B \}}}{T})}\bigg]-\phi_{B\{\bar{B}\}}(n_{B\{\bar{B}\}})
\end{equation}
    
\begin{equation}
P_M(T)=-T\sum_{i\in M}\int d\Gamma_{i} \ \text{ln}\bigg[1- e^{-(\frac{\varepsilon_i}{T})}\bigg]-\phi_M(n_M)
\end{equation}

The energy densities for baryons and mesons are given by

\begin{equation}
\epsilon_{B\{\bar{B}\}}(T,\mu_B)=\sum_{i\in B\{\bar{B}\}}\int d\Gamma_{i}\frac{\varepsilon_i} {\bigg[ e^{(\frac{E_i-\mu_{\text{eff},
B\{\bar B\} } } {T})}+1\bigg]} +\phi_{B\{\bar{B}\}}(n_{B\{\bar{B}\}})
\end{equation}

\begin{equation}
\epsilon_M(T)=\sum_{i\in M}\int d\Gamma_{i}\frac{\varepsilon_i}{\bigg[e^{\frac{\varepsilon_i}{T}}-1\bigg]}+\phi_M(n_M)
\end{equation}

\section{Results}
\label{secIII}

	Let us now discuss the results that we have obtained for specific heat ($C_V$),  isothermal compressibility ($\kappa_T$) 
	and speed of sound ($C_s^2$) as a function of temperature ($T$) at fixed values of baryon chemical potential ($\mu_B$). We shall take two representative values of baryon chemical potential:  $\mu_B=0$ and $0.3$ GeV. We have chosen these values of $\mu_B$ because the critical end point is expected to be near $\mu_B$=0.3 GeV~\cite{Stephanov:1998dy,Parotto:2018pwx}.  The only parameters in our model are $K_M$ and $K_B$.  We choose three different representative values for meson mean field parameter, $viz.$, $K_M=0, 0.1$ and $0.15$ GeV$\cdot$fm$^{3}$,  while we fix baryon mean-field parameter $K_B=0.45$ GeV$\cdot$fm$^{3}$. It has been found that  this choice of parameters leads to good agreement with LQCD data for pressure and some of the cumulants of baryon number susceptibilities~\cite{Huovinen:2017ogf,Pal:2020ucy,Kadam:2019peo}. For comparison with excluded-volume HRG model (EVHRG),  we have taken radii of hadrons as $R_M=0$ (for mesons) and  $R_B=0.3 \:\text{fm}$ (for baryons and anti-baryons). This choice is commensurate with the study of Ref.~\cite{Vovchenko:2017xad}. Here we have assumed repulsive interaction among baryon-baryon and antibaryon-antibaryon pairs while there is no repulsive interaction among other pairs. Consideration of the 
absence of repulsive interaction among meson pairs and presence of repulsive interaction among baryon pairs and anti-baryon pairs, separately, lead to a better agreement of baryon number susceptibilities with Lattice data. This assumption is consistent with Refs.~\cite{Vovchenko:2016rkn,Satarov:2016peb}. In this work we have taken all the hadrons and resonances up to 2 GeV listed in particle data review~\cite{Tanabashi:2018oca}.

	

		In case of ideal HRG model expression for the specific heat is written as
		\begin{eqnarray}
	C_{V,\mu_B}(T)&=&\bigg(\frac{\partial \epsilon}{\partial T}\bigg)_{V,\mu_{B}}\\\nonumber
	&=&\sum_i \int_{0}^{\infty}d\Gamma_i \frac{ E_i(E_i-B_i\mu_B)\exp[(E_i-B_i\mu_B)/T]}{T^2\{\exp[(E_i-B_i\mu_B)/T]\pm 1\}^2}\\\nonumber
	&=&\sum_i\int_{0}^{\infty} d\Gamma_i \: f_{\pm}(1\mp f_{\pm})\frac{E_i(E_i-B_i\mu_B)}{T^2}
	\end{eqnarray}
        where $f_{\pm}$ are the distribution functions defined as
	\begin{equation}
	f_{\pm}=\frac{1}{e^{(x/T)}\pm1}
	\label{df}
	\end{equation}
	with $x=E_i-B_i\mu_B$ for pure HRG model. 
	
	In case of HRGMF model, the expression for the baryonic contribution to the specific heat picks up mean-field dependent terms. For baryons, with $x=E_i+K_Bn_B-B_i\mu_B$, we get

		\begin{eqnarray}
	C_{V,\mu_B}(T)=\sum_{i \in B}\int d\Gamma_i\: f^{\text{mf}}_{+}(1- f^{\text{mf}}_{+})(E_i+K_Bn_B) \nonumber \\
	\times \Bigg[\frac{(E_i+K_Bn_B-B_i\mu_B)}{T^2} 
	-\frac{K_B}{T}\Bigg(\frac{\partial n_B}{\partial T}\Bigg )_{V,\mu_B}\Bigg]\\ \nonumber
	=\sum_{i \in B}\int d\Gamma_i\: f^{\text{mf}}_{+}(1- f^{\text{mf}}_{+})\varepsilon_{i} 
	\Bigg[\frac{x}{T^2}
	-\frac{K_B}{T}\Bigg(\frac{\partial n_B}{\partial T}\Bigg )_{V,\mu_B}\Bigg]
	\end{eqnarray}
	Where 
	\begin{equation}
		\frac{\partial n_{B}}{\partial T}=\sum_{i\in B}\frac{\int d\:\Gamma_i\:\frac{x}{T^2}f_{+}^{\text{mf}}(1- f_{+}^{\text{mf}})}{1+\frac{K_B}{T}\int d\Gamma_{i}\: f^{\text{mf}}_{+}(1- f_{+}^{\text{mf}})}
		\end{equation}
		Similar expressions can be written for antibaryons with $n_{\bar B}$  and mesons  with $f_-^{\text{mf}}, \ K_M$ and $n_M$.  The form of the distribution functions in presence of mean-field interactions ($f^{\text{mf}}_{\pm}$) are 
		the same as those without interactions given in equation~(\ref{df}). However the argument 
		$x$ changes as $x= E_i+K_Bn_B-B_i\mu_B$ for baryons and antibaryons and $x= E_i+K_Mn_M$ for 
		mesons. 
		
	\begin{figure}[t]
				\hspace{-1.2cm}
			\begin{tabular}{c c c}
		\includegraphics[scale=0.6]{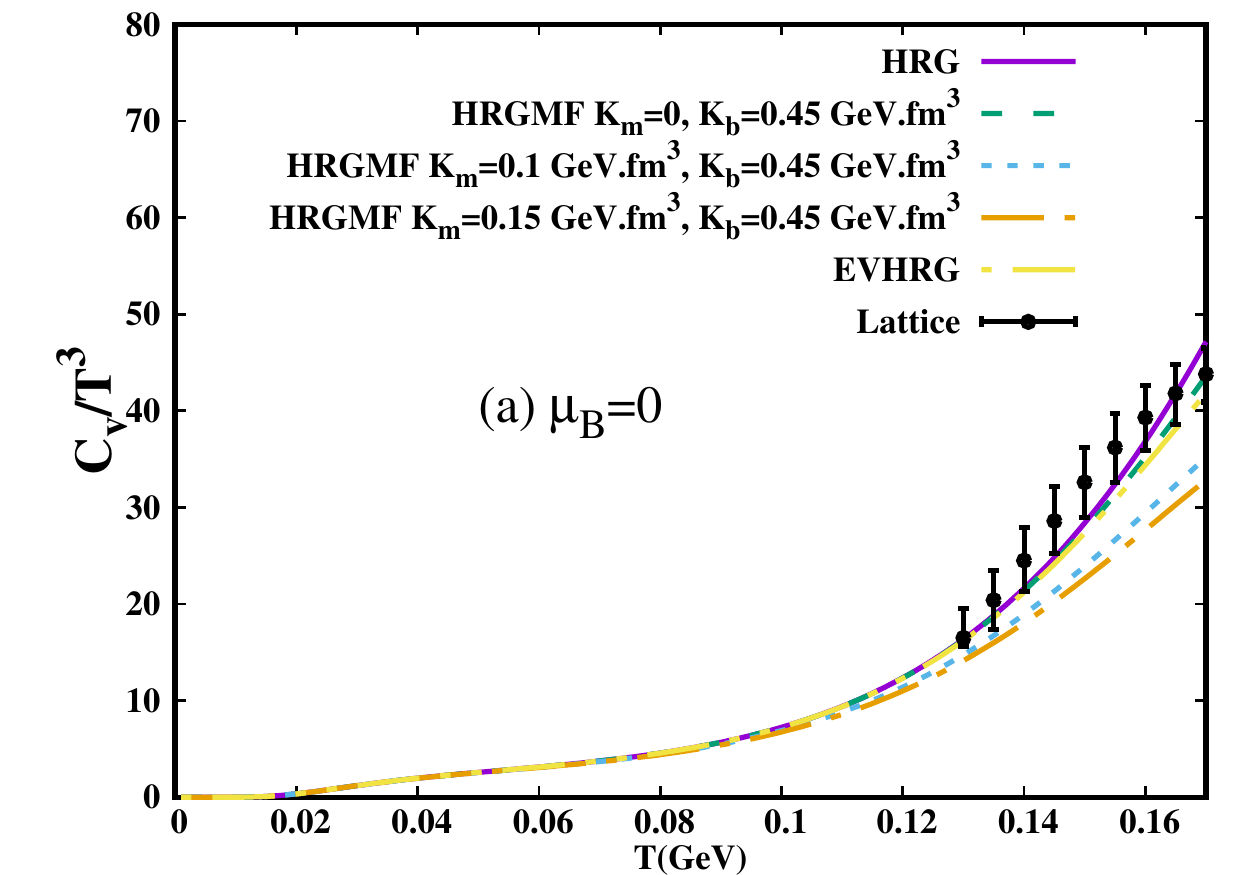}  
		\hspace{-0.4cm}
		\includegraphics[scale=0.6]{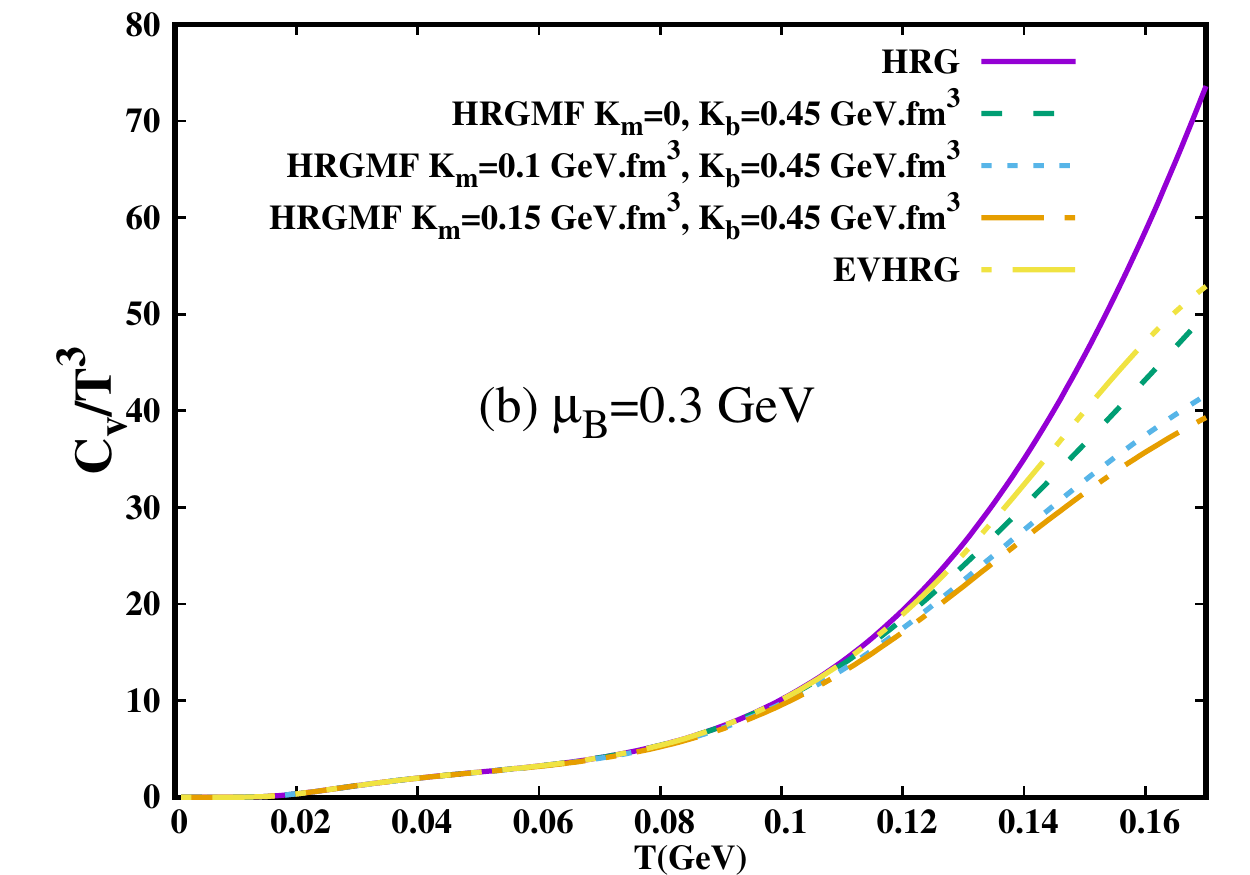} \\
		\hspace{-0.4cm}
		\includegraphics[scale=0.6]{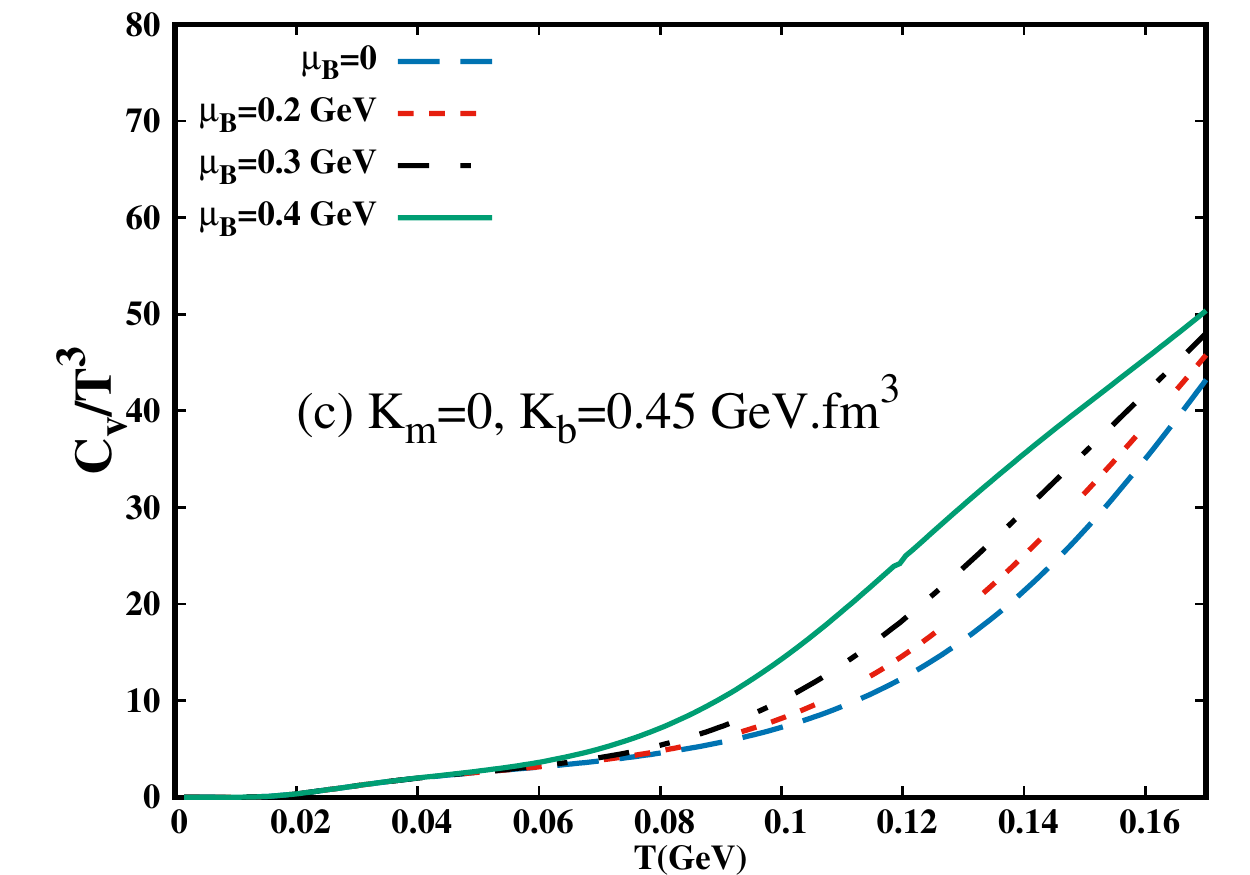}
	
			\end{tabular}
		\caption{(Color Online) Scaled specific heat as a function of temperature. The first two panels are for $\mu_B = 0$ and 0.3 GeV respectively. Third panel shows scaled specific heat for $K_m=0$ at $\mu_B = 0,$ 0.2, 0.3 and 0.4 GeV.}
		\label{cv}
	\end{figure}

	 Figs.~\ref{cv}(a)  and \ref{cv}(b) show $C_V/T^3$ as a function of temperature for two different values of $\mu_B$ i.e. 0 and 0.3 GeV for ideal HRG, EVHRG and HRG mean field model (HRGMF).   (In this work the notation $C_V$ means specific heat 
	 calculated at constant volume and constant $\mu_B$).  We note that $C_V$ increases monotonically with temperature for ideal HRG (solid purple curve).  This rise is significantly rapid near $T \sim 0.15$ GeV which is close to $T_c$ where the phase transition from hadronic phase to Quark-gluon plasma (QGP) phase is expected to occur as suggested by lattice QCD simulations. In fact, near the QCD critical point the specific heat is expected to show a power law behaviour $C_V\sim |t|^{-\alpha}$, where $t=\frac{T-T_c}{T_c}$ and $\alpha$ is the critical exponent.  However,  in the hadron resonance gas model we only observe rapid rise as temperature reaches $T_c$.  In case of HRG, with discrete mass spectrum and with only spin degree of freedom $J_i$, the spin degeneracy $(2J_i+1)$ results in hadron resonance abundances which goes like $g_i \propto m_i^2 $.  Thus the partition function and all of its higher order derivatives are continuous.  Hence, the specific heat does not diverge as $T\longrightarrow T_c$. The effect of repulsive  interactions is to suppress the rapid rise of $C_V$ at high temperature as can be noted in Fig.\ref{cv}. This suppression is not hard to understand. By definition, $C_V$ is the amount of energy required to raise the temperature by unit value. If there is a repulsive interaction between hadrons then the energy required to create a given species of hadron in a medium is increased.  Hence the corresponding yield will be reduced.  Thus, the energy poured into the system is utilised to raise the temperature rather than to create massive hadrons.  Upshot of this is that the contribution of degrees of freedom, which are hadrons in case of hadronic matter, to the  scaled specific heat $C_V/T^3$,  will be suppressed in the models with repulsive interactions. 
	
	 For $\mu_B = 0$, we have confronted our results with that obtained in Lattice QCD~\cite{Bazavov:2014pvz}. It is clearly seen that at temperatures higher than $0.15 $ GeV HRGMF results are in good agreement with the lattice data. The inclusion of mesonic mean-field interactions slightly underestimates the lattice data while reproducing the overall behaviour. The EVHRG result lies close to HRGMF result for $K_M=0$ as we have not considered repulsive interaction among meson pairs.  Although confronting the HRG model and its extension beyond $T_c$ may not be reasonable, our study clearly  indicate that the repulsive interactions play an important role near $T_c$.
	 
	Fig.~\ref{cv}(c)  shows specific heat estimations of HRGMF model with $K_M=0$ and $K_B=0.45$ GeV $\text{fm}^3$  at various values of $\mu_B$. We note that the scaled ratio $C_V/T^3$ is higher in magnitude at higher $\mu_B$. 

	Isothermal compressibility ($\kappa_T$) is defined as
	\begin{equation}
	\kappa_T|_{T,\langle N_i\rangle}=-\frac{1}{V}\bigg(\frac{\partial V}{\partial P}\bigg)_{T,\langle N_i\rangle}
	\label{kt0}
	\end{equation}
	Considering the pressure as a function of temperature and individual chemical potentials $\mu_i$ of the particle species,
	\begin{equation}
	dP=\Big( \frac{\partial P}{\partial T}\Big)dT+\sum_i \Big(\frac{\partial P}{\partial \mu_i}\Big)d\mu_i
	\end{equation}
    From which we get
	\begin{equation}
    \Big( \frac{\partial P}{\partial V}\Big)_{T,\langle N_i\rangle}=\sum_i \Big(\frac{\partial P}{\partial \mu_i}\Big)\Big( \frac{\partial \mu_i}{\partial V}\Big )\Big \vert_{T,\langle N_i\rangle}
    \label{delpv}
	\end{equation}
	The change in the number of particle species $N_i$ is given by
	\begin{equation}
	dN_i=\Big( \frac{\partial N_i}{\partial T}\Big)dT+\Big( \frac{\partial N_i}{\partial \mu_i}\Big)d\mu_i+\Big( \frac{\partial N_i}{\partial V}\Big)dV
	\end{equation}
	The above equation, for constant $N_i$ and T, reduces to 
	\begin{equation}
	\Big( \frac{\partial \mu_i}{\partial V}\Big )\Big \vert_{T,\langle N_i\rangle}=-\Big( \frac{\partial N_i}{\partial V}\Big)\Big/\Big( \frac{\partial N_i}{\partial \mu_i}\Big)
	\end{equation}
	Using $\frac{\partial N}{\partial V}=\frac{\partial P}{\partial \mu}$ we have from eq.~(\ref{delpv})
	\begin{equation}
	 \Big( \frac{\partial P}{\partial V}\Big)_{T,\langle N_i\rangle}=-\sum_i\frac{\Big(\frac{\partial P}{\partial \mu_i}\Big)^2}{\Big( \frac{\partial N_i}{\partial \mu_i}\Big)}
	 \label{kt1}
	\end{equation}
	Using equations~(\ref{kt0}) and~(\ref{kt1}),
	\begin{equation}
		\kappa_T|_{T,\langle N_i\rangle}=\frac{1}{\sum_i[(\frac{\partial P}{\partial\mu_i})^2/{(\frac{\partial n_i}{\partial\mu_i})]}}
	\end{equation}
    It is to be noted that here $\mu_i$ is not  the equilibrium chemical potential related to the whole system. Instead, it is the individual chemical potential linked to the particle species 'i'. $\mu_i$ is related to $\mu_B$, $\mu_Q$, and $\mu_S$ by the relation $\mu_i=B_i\mu_B+Q_i\mu_Q+S_i\mu_S$. $n_i$ appearing in this expression is not net density but the density of $i^{th}$ species which is not zero. So in evaluating the sum one first writes the expression for P and $n_i$ as a function of $\mu_i$ and takes the derivative w.r.t. $\mu_i$ defined for that species. Hence $\kappa_T$ gets contribution from all the species: mesons as well as baryons. The term  $\frac{\partial n_i}{\partial \mu_i}$  for a given baryon or meson species in HRG is given by
	
	\begin{equation}
	\frac{\partial n_i}{\partial \mu_i}=\int_{0}^{\infty}d\Gamma_{i}\:  \frac{1}{T}\:f_{\pm}[1\mp f_{\pm}]	
	\end{equation}
	
	while, in the HRGMF model it is given by,
	\begin{equation}
	\frac{\partial n_i}{\partial \mu_i}=\int_{0}^{\infty}d\Gamma_{i}\:  \frac{1}{T}\:f^{\text{mf}}_{\pm}[1\mp f^{\text{mf}}_{\pm}]	\bigg(1-K_{B\{M\}}\frac{\partial n_{B\{M\}}}{\partial \mu_i}\bigg)
	\end{equation}
	Here the upper sign is for baryons and lower sign is for mesons. $K_B$ and $K_M$ are used for baryons and mesons respectively. (Here we have worked with 
	charge chemical potential ($\mu_Q$) and strangeness chemical potential ($\mu_S$) to be zero). 

	\begin{figure}[t]
		\hspace{-1.2cm}
					\begin{tabular}{c c c}
		\includegraphics[scale=0.6]{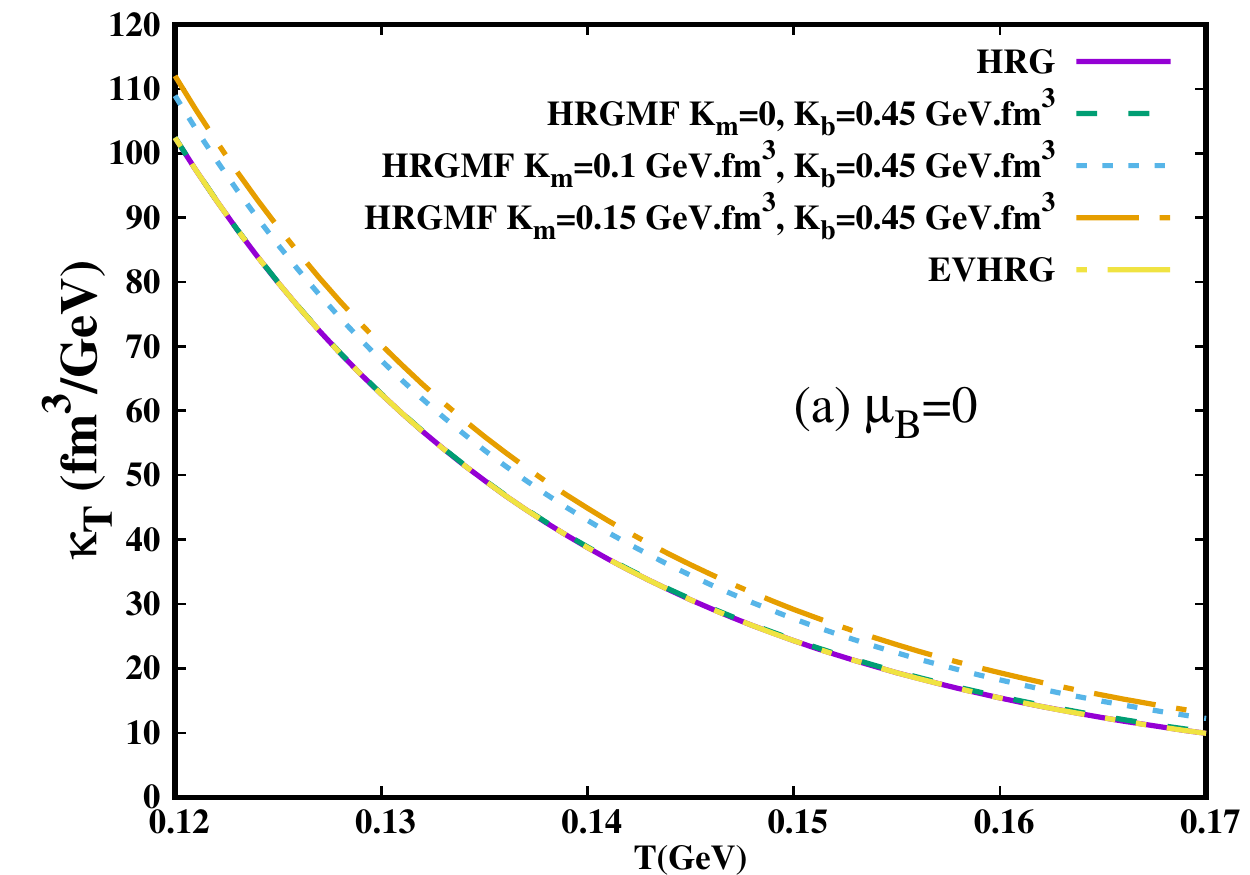} 
		\hspace{-0.4cm}
		\includegraphics[scale=0.6]{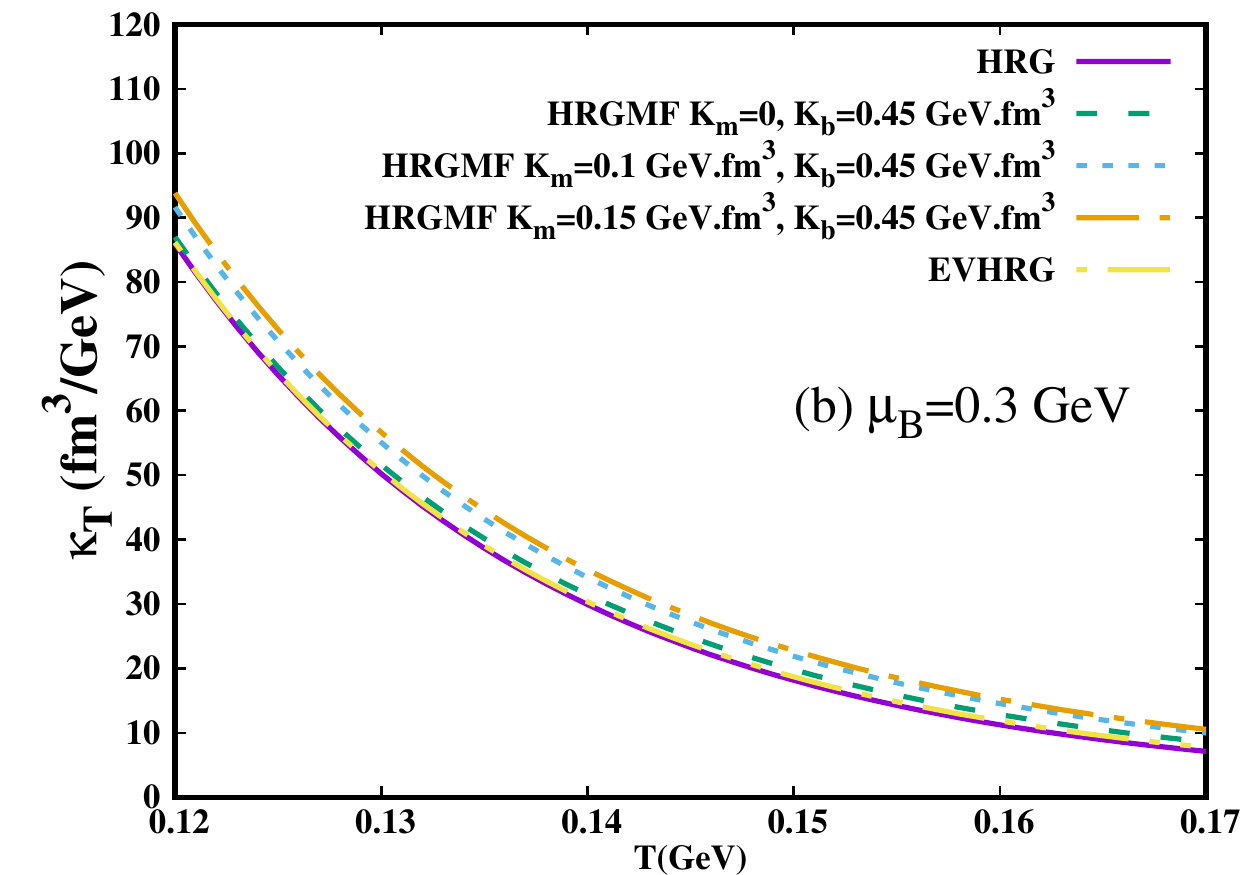} \\
		\hspace{-0.4cm}
		\includegraphics[scale=0.6]{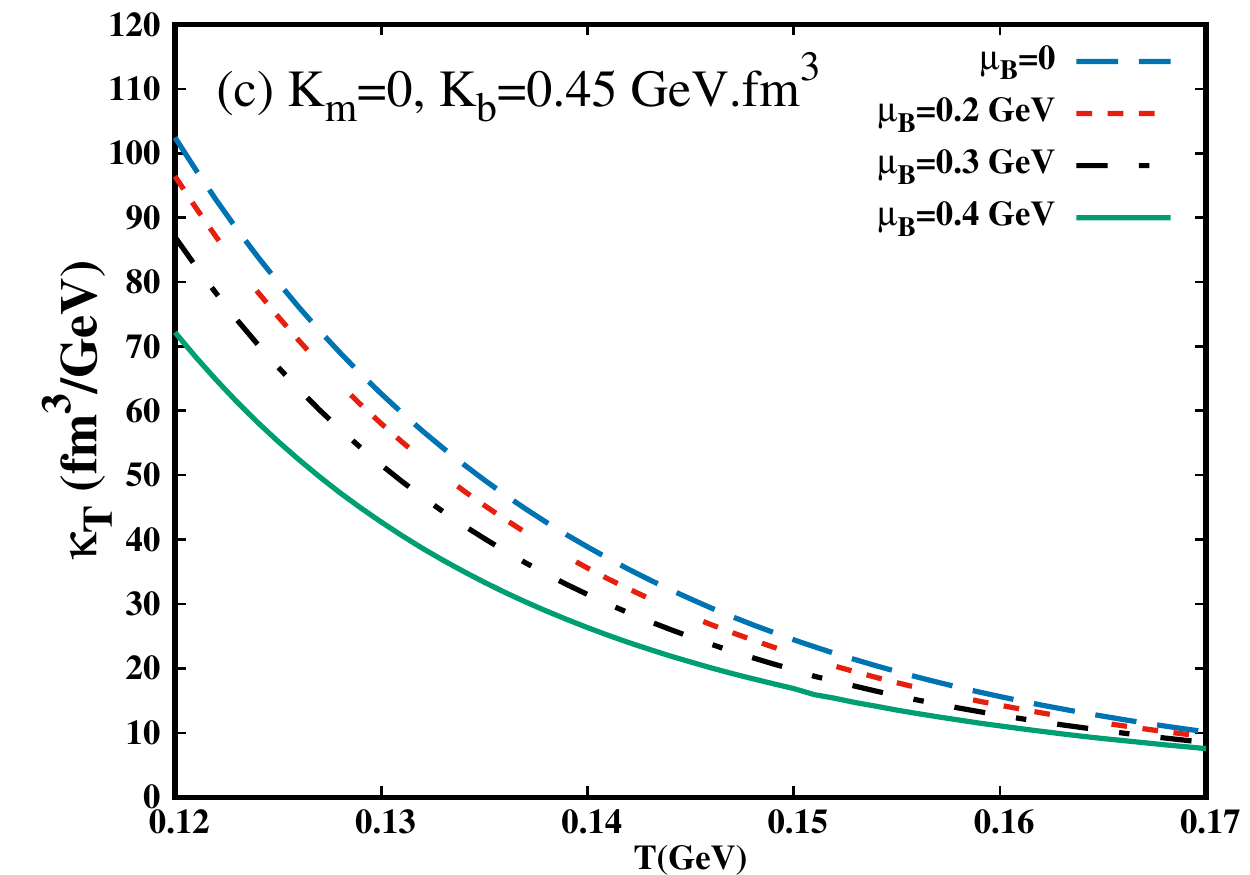}
		\end{tabular}
			
		\caption{(Color Online) Isothermal compressibility as a function of temperature. The first two panels are for $\mu_B = 0$ and 0.3 GeV respectively. Third panel shows compressibility for $K_m=0$ at $\mu_B = 0,$ 0.2, 0.3 and 0.4 GeV.}
				\label{kt}
	\end{figure}

	 Figs.~\ref{kt}(a) and ~\ref{kt}(b)  show plots of isothermal compressibility $\kappa_T$ as a function of temperature at baryon chemical potential 0 and 0.3 GeV respectively. At $\mu_B=0$ (Fig.\ref{kt}(a)), for all the three variants of HRG model considered, $\kappa_T$ decreases with increasing temperature. In case of HRG, this decreasing behaviour of $\kappa_T$ as a function of $T$ is expected as $\kappa_T\propto \text{pressure}^{-1}$.  When the temperature is increased, more and more hadrons populate the system and hence the pressure also increases which render the decreasing behaviour of $\kappa_T$. In case of HRGMF,  the scaled pressure ($P/T^4$) is negligibly smaller compared to ideal HRG pressure. The choice of excluded volumes in this work ($R_M=0$ for mesons and $R_B$=0.3 fm for baryons and anti-baryons) lead to only small deviation in compressibility as compared to HRG results, which, for $\mu_B=0$, is negligible. The isothermal compressibility is more sensitive to the sizes of the mesons rather than those of the baryons. This is corroborated by the fact that in the HRGMF model, for $K_M=0$, the result is almost same as ideal HRG model. So the repulsion in the mesonic sector seems to contribute more compared to that in the baryonic sector. At finite $\mu_B$ (Fig.\ref{kt}(b)), while the overall behaviour of $\kappa_T$ as a function of $T$ is similar to the $\mu_B=0$ case,  the magnitude of $\kappa_T$ is smaller than that at $\mu_B=0$ case. This behaviour is the reflection of that fact that the scaled pressure ($P/T^4$) is higher at finite $\mu_B$ as compared to that at $\mu_B=0$.

	The (adiabatic) speed of sound is defined as 
	\begin{equation}
	C_s^2=\bigg(\frac{\partial P}{\partial \epsilon}\bigg)_{s/n}
	\end{equation}

	Physical waves propagate at constant $s/n$. In order to discuss the implications of our results in the context of HIC, which we shall do in next section, we need to find explicit expression for adiabatic speed of sound which is also valid at finite $\mu_B$. It can be easily shown that at constant $s/n$ the speed of sound is~\cite{Albright:2015fpa}
	
	\begin{equation}
	C_s^2(T,\mu_B)=\frac{v_n^2 Ts+v_s^2\mu_Bn_B}{\epsilon+P}
	\end{equation}
	
	where,
	
	\begin{equation}
	v_s^2(T,\mu_B)=\frac{n_B\chi_{TT}-s\chi_{\mu_B T}}{\mu_{B}(\chi_{TT}\chi_{\mu_B\mu_B}-\chi_{\mu_BT})}
	\end{equation}
	is the speed of sound at constant $s$ and,
	
	\begin{equation}
	v_n^2(T,\mu_B)=\frac{s\chi_{\mu_B\mu_B}-n_B\chi_{\mu_B T}}{T(\chi_{TT}\chi_{\mu_B\mu_B}-\chi_{\mu_BT})}
	\end{equation}
	is the speed of sound at constant $n$. The susceptibilities ($\chi$'s) are defined as, $\chi_{ab}=\partial^2 P/\partial a \partial b $, with $a,b=T,\mu_B$.

	\begin{figure}[t]
		\hspace{-1.2cm}
					\begin{tabular}{c c c}
		\includegraphics[scale=0.6]{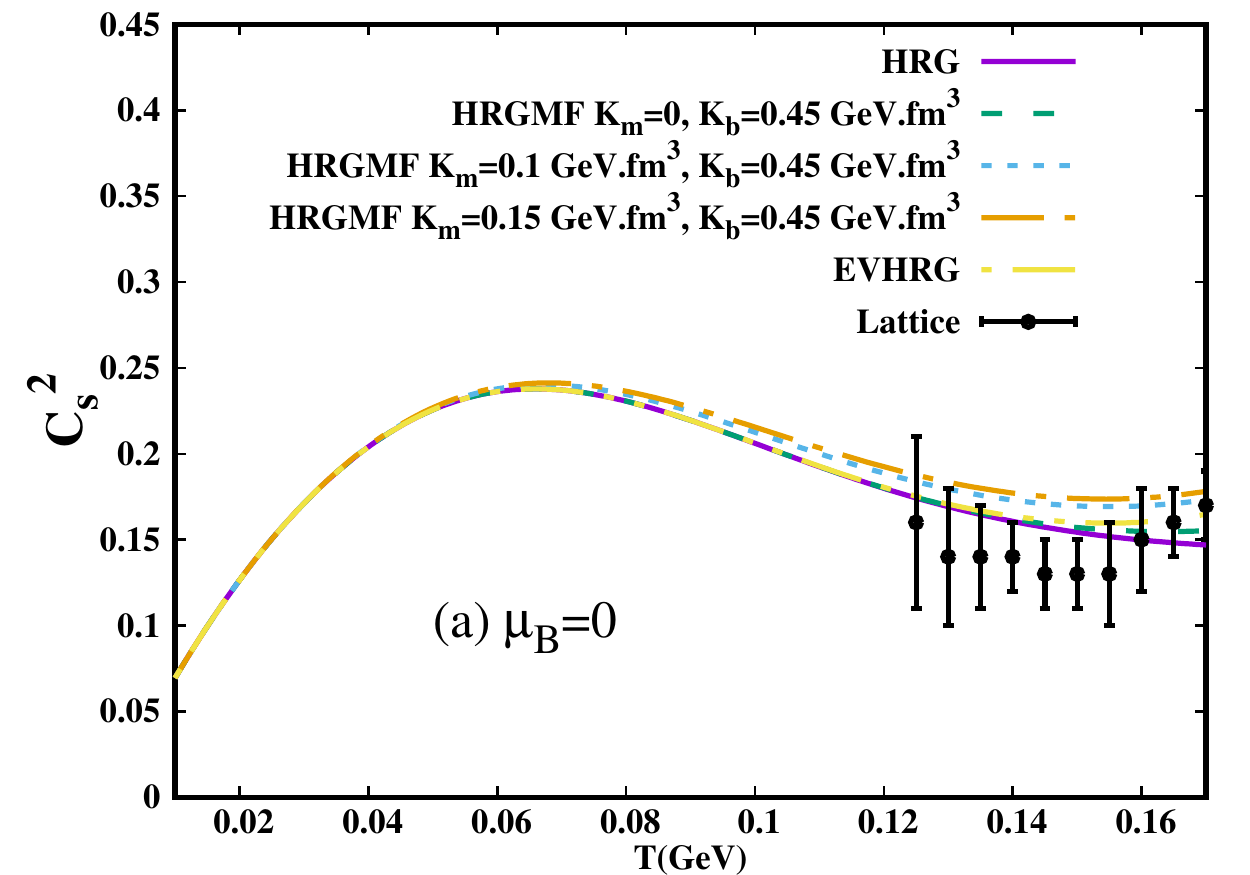}
		\hspace{-0.4cm}
		\includegraphics[scale=0.6]{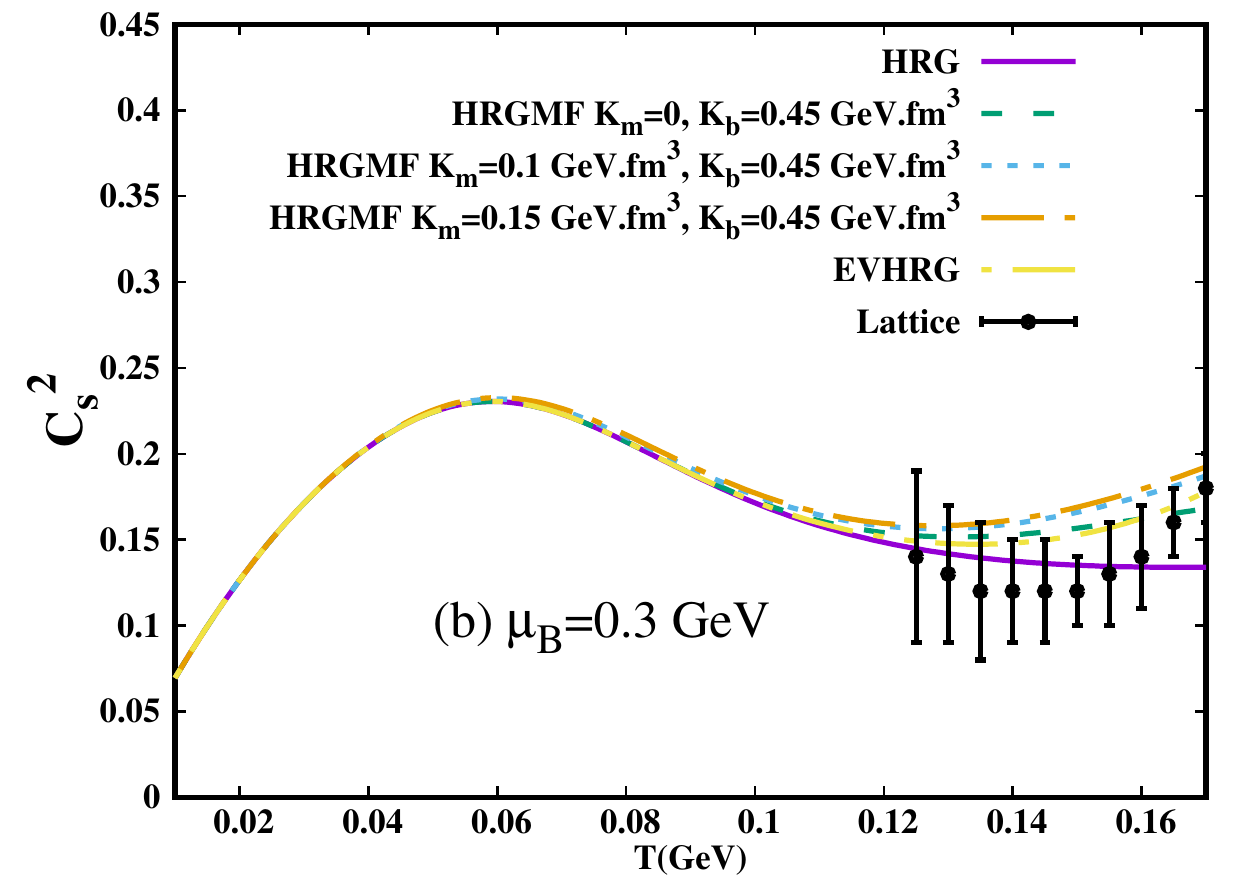} \\
		\hspace{-0.4cm}
		\includegraphics[scale=0.6]{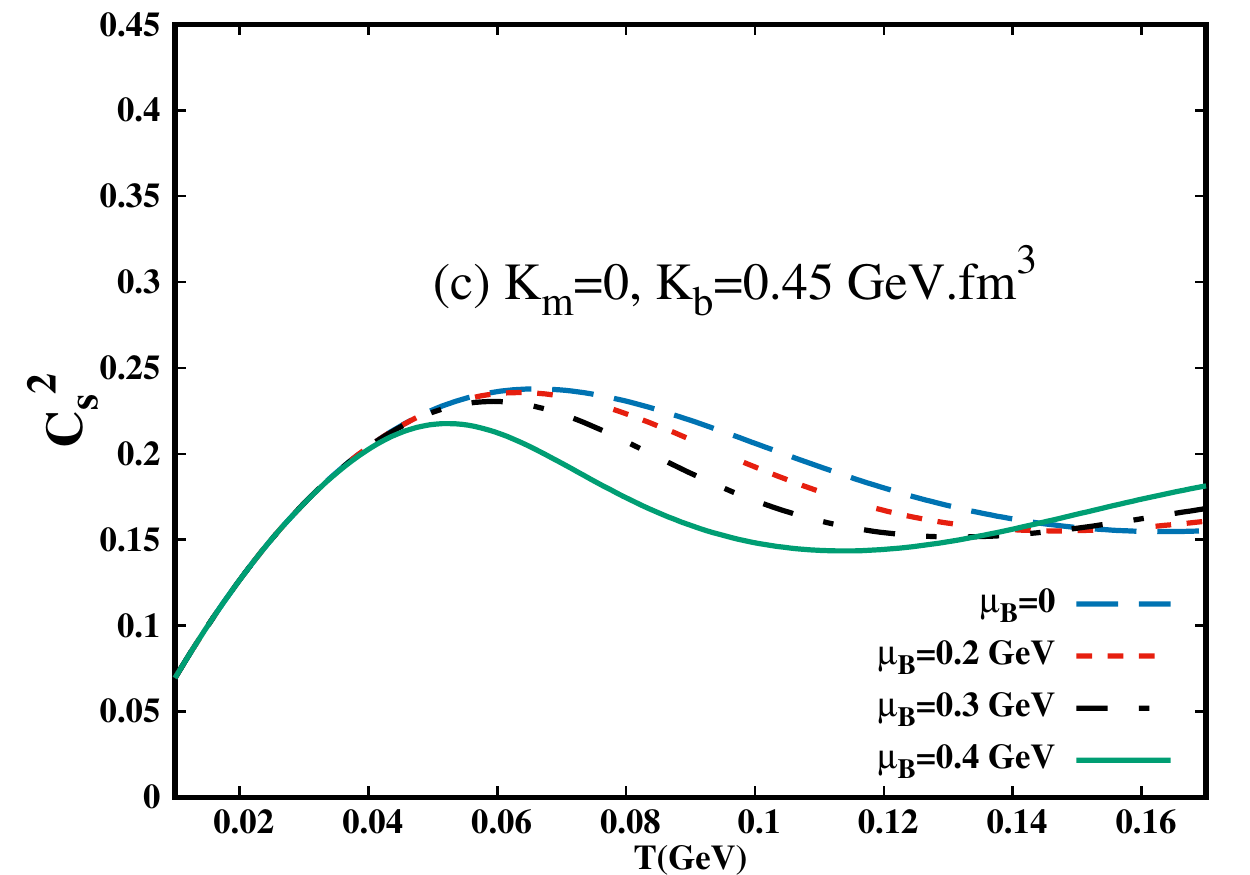}
		\end{tabular}
	
		\caption{(Color Online) Speed of sound squared ($C_s^2$) as a function of temperature. The first two panels are for $\mu_B = 0$ and 0.3 GeV respectively. Third panel shows speed of sound squared ($C_s^2$) for $K_m=0$ at $\mu_B = 0,$ 0.2, 0.3 and 0.4 GeV.}
			\label{cs}
	\end{figure}

	In Fig.~\ref{cs} we have shown the variation of $C_s^2$  with temperature for different values of baryon chemical potential.  The $C_s^2$ depends on the degrees of freedom of the system, the equation of state and hence the interaction among the hadrons.  At very low temperatures $T\sim m_\pi$, energy, pumped into the system, is utilised to 
	increase the temperature of matter which is dominated mostly by lighter pions. Hence the change in pressure is dominant over 
	change in energy density of the system resulting in monotonic rise in speed of sound.   This can be observed even with the 
	repulsive interactions and at $\mu_B\neq 0$. As the temperature increases more massive hadrons and hadronic resonances start
	to populate the system.  The energy pumped into the system is utilised to create heavier hadrons.  The upshot is that increase
	in pressure lags behind the corresponding increase in the energy density.  Thus, the rapid rise of $C_s^2$ slows down and 
	attains maximum around $T\sim 0.08$ GeV.  After that it decreases slowly and saturates at a constant value around $T\sim 
	0.16$ GeV.  The effect of repulsive interaction does not show up until the system achieves an energy density sufficient to 
	create baryons.  Again,  because in presence of repulsive interactions it will take more energy to create a baryon,  increase in 
	energy density lags behind increase in pressure.  This is reflected in Fig.\ref{cs} where we note that the magnitude of $C_s^2$
	is greater in HRGMF and EVHRG models as compared to HRG model.  Also, with repulsive interactions turned on, $C_s^2$
	increases with temperature. However, with medium dependent hadron masses the speed of sound estimated within EVHRG model does not rise rapidly at high temperature and it has been shown to be in good agreement with lattice QCD data~\cite{Kadam:2015dda}. An interesting observation with regard to the minimum in $C_s^2$ can be noted from Fig.(\ref{cs}). The minimum of the speed of sound, in the case of a system with repulsive interactions, is shifted to lower temperature as compared to ideal gas. We compare our results with those obtained with lattice QCD~\cite{Borsanyi:2012cr}. As in the case of specific heat, our results with repulsive interaction, at temperatures higher than $0.15 \ GeV$, approaches the lattice results. For all the values of $\mu_B$ we have an excellent fit with the lattice results. Furthermore, as the temperature increases, more and more repulsion seems to be preferable.  Fig.~\ref{cs}(c) shows speed of sound at $K_M=0$ for various values of $\mu_B$. There is almost no difference in $C_s^2$ among various $\mu_B$ cases at low temperatures, since, number of baryons is smaller at low temperatures. At mid-range temperatures, $C_s^2$ decreases as $\mu_B$ is increased. This is because the increase in pressure lags behind the increase in energy density  more for higher $\mu_B$. The trend becomes opposite at high temperatures when the increase in energy density lags behind the increase in pressure again.

	\section{Discussion}
	\label{secIV}
	
	Let us now discuss the implications of our results in the light of heavy-ion collision experiments.  Connection of any physical quantity with the heavy-ion collision experiments can be made by finding the beam energy ($\sqrt{s}$) dependence of the temperature and chemical potential. This is extracted from a statistical thermal model description of the particle yield at various $\sqrt{s}$~\cite{Cleymans:2005xv}. The freeze-out curve is parametrised by 
	
	\begin{equation}
	T(\mu)=a-b\mu^2-c\mu^4;\hspace{0.1 cm} \mu=d/(1+e\sqrt{s})
	\label{FO}
	\end{equation}	
	where $a=0.140$ GeV, $b=139\pm 0.016\:\text{GeV}^{-1}$, $c=0.053\pm0.021\:\text{GeV}^{-3}$, $d=1.308\pm 0.028$ GeV and $e=0.273 \pm 0.008\:\text{GeV}^{-1}$. Note that this parametrization is based on ideal HRG model.

	For a system in equilibrium, the event-by-event  temperature fluctuation is controlled by  heat capacity 
	\begin{equation}
	P(T)\sim \text{exp}\bigg (-\frac{{\tilde C_{V,N}}}{2}\frac{(\Delta T)^2}{\langle T \rangle^2}\bigg)
	\end{equation}
	where $\Delta T=T-\langle T \rangle$ and $\langle T \rangle$ is the mean temperature. This yields the relation between  heat capacity $(\tilde C)$ and temperature fluctuation~\cite{LL}  as 
	\begin{equation}
	\frac{1}{\tilde{C}_{V,N}}=\frac{\langle T^2 \rangle - \langle T \rangle^2}{\langle T \rangle^2}
	\label{sh}
	\end{equation}
	In the above equation, the heat capacity is calculated at constant $(V,N)$. On the other hand, we have 
	calculated the specific heat at constant $(V,\mu)$.  The relationship between $C_{V,N}$ and $C_{V,\mu}$ is given by
	\begin{equation}
		C_{V,N}=C_{V,\mu}-T\frac{{\big(\frac{\partial s}{\partial \mu}}\big)_{T,V}{\big(\frac{\partial n}{dT}\big)}_{\mu,V}}{{\big(\frac{\partial n}{\partial \mu}\big)}_{T,V}}
	\end{equation}
	The equation~(\ref{sh}) is valid in the presence of a large energy reservoir. This role may be played by the longitudinal degrees of freedom of the beam in heavy-ion collision~\cite{Stodolsky:1995ds}. Thus, the specific heat can actually be extracted using temperature fluctuations through  $p_T$ spectra measured on the basis of  event-by-event analysis of experimental data. While such analysis could be done, we restrict ourselves only to the beam-energy dependence of $C_V$ extracted using  Eq.(\ref{FO}). This will give information about the magnitude of specific heat at the freeze-out.
	
\begin{figure}[h]
	\hspace{-1.2cm}
					\begin{tabular}{c c}
		\includegraphics[scale=0.6]{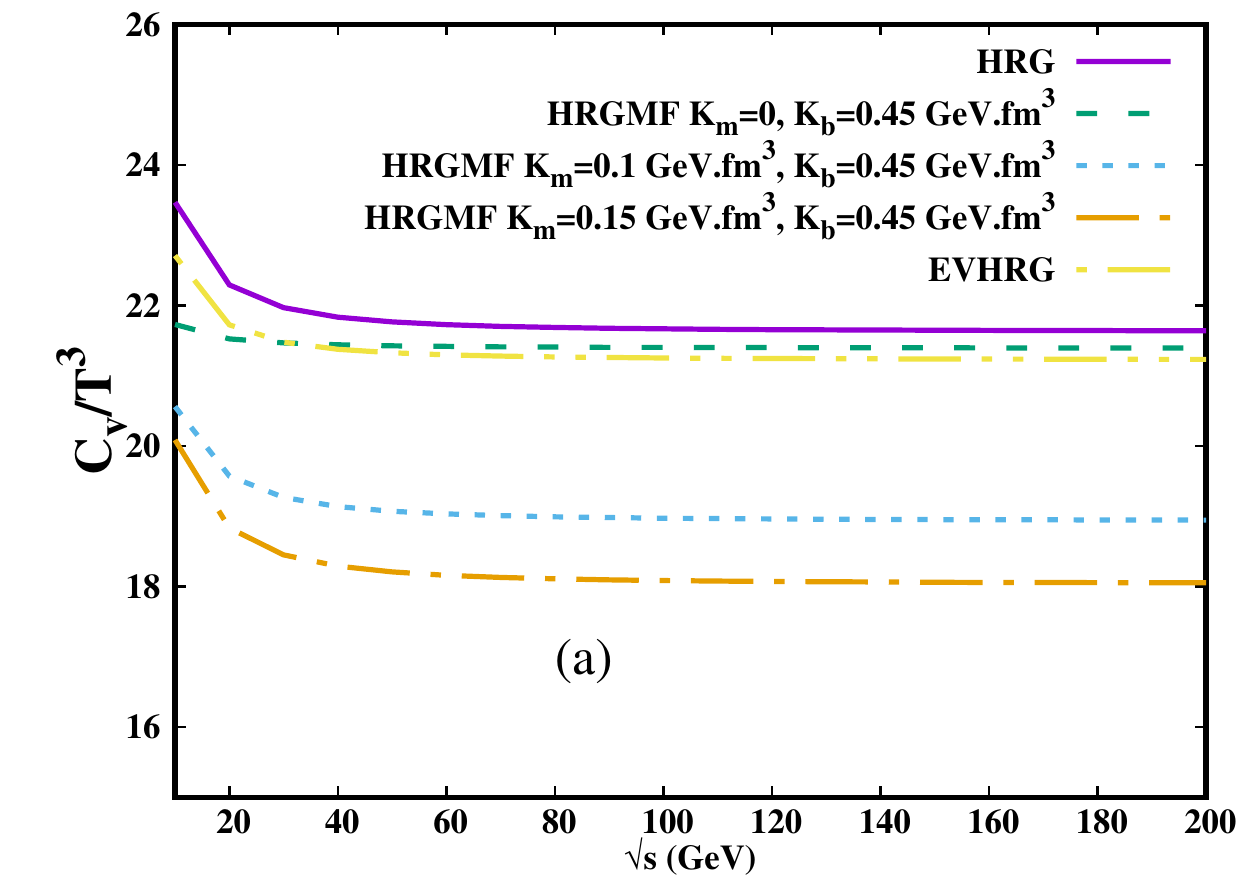}
		\hspace{-0.4cm}
		\includegraphics[scale=0.6]{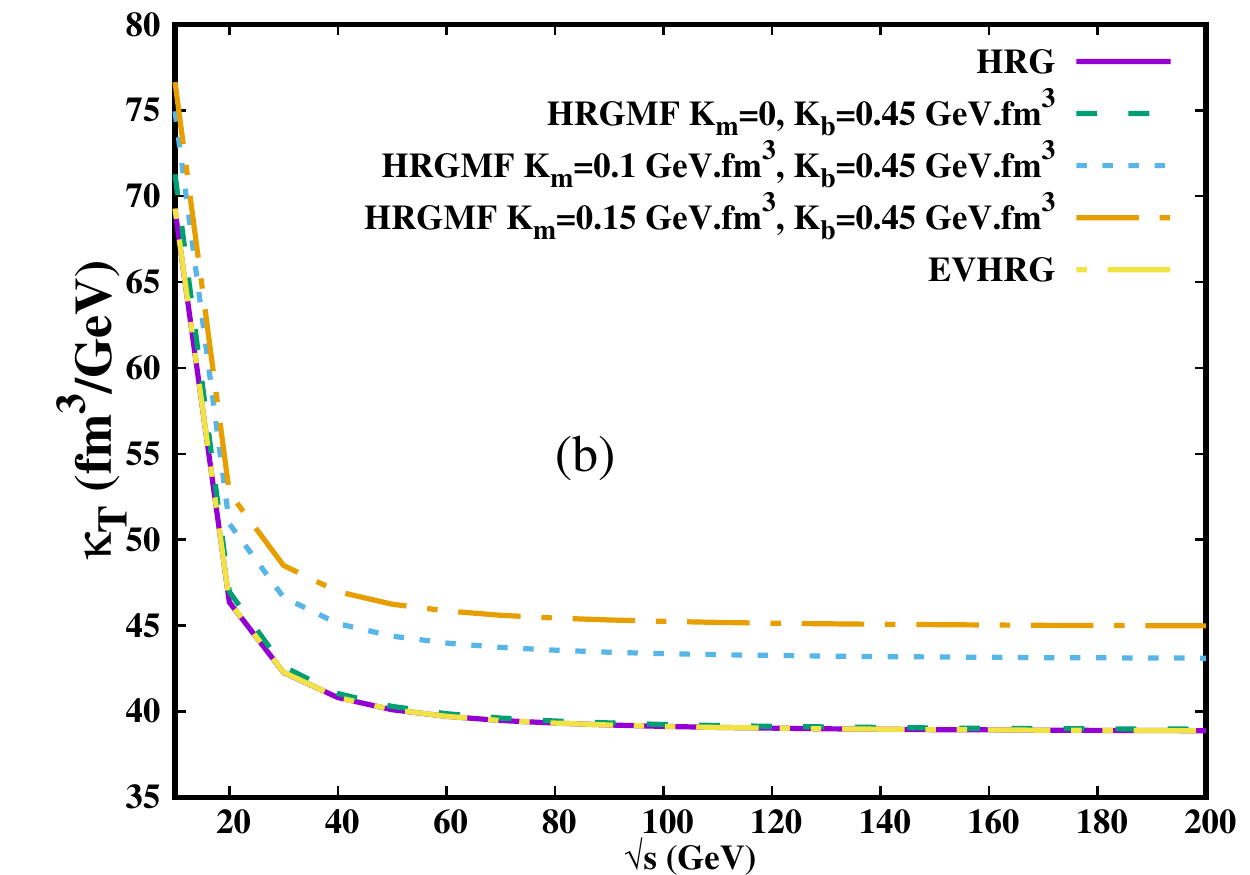} \\
		\hspace{-0.4cm}
		\includegraphics[scale=0.6]{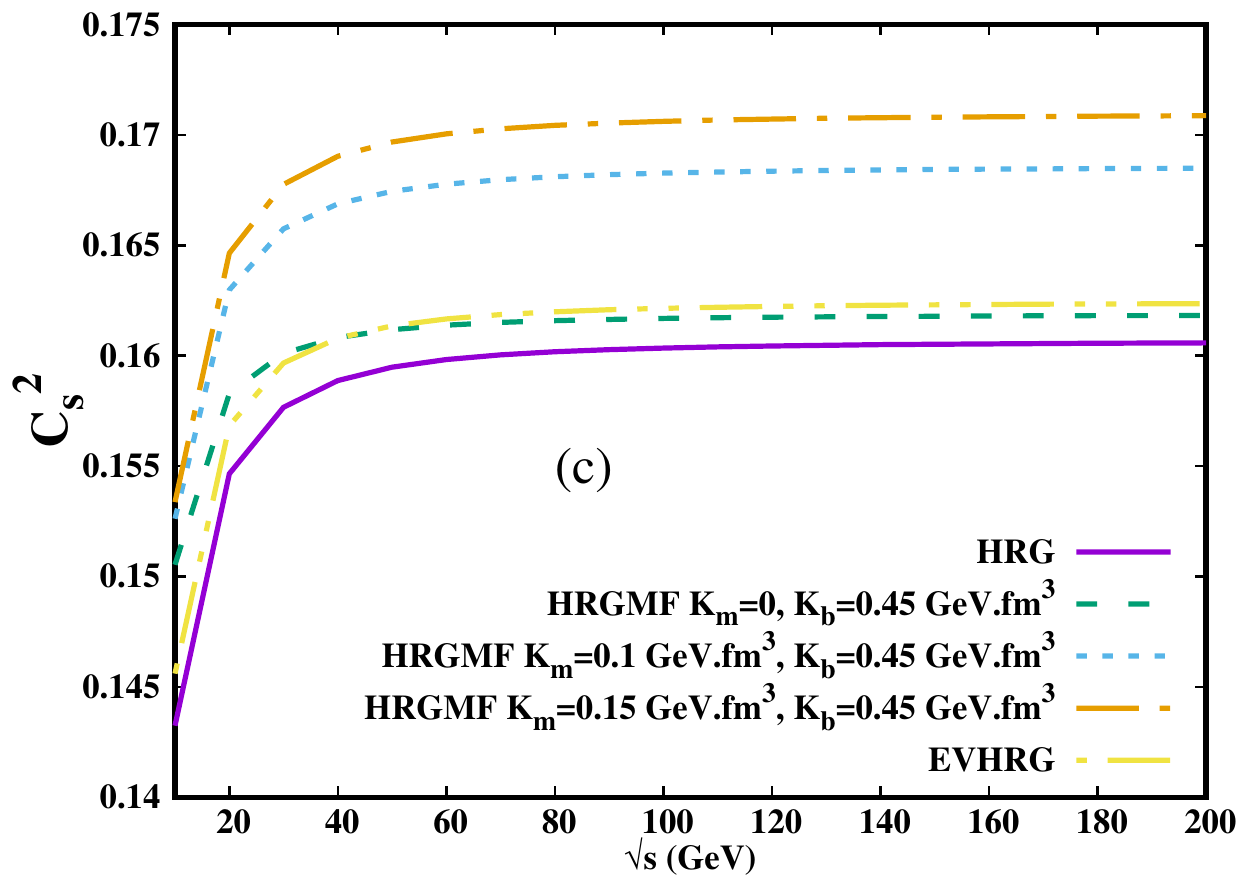}
		\end{tabular}
				
		\caption{(Color Online) Collsion energy dependence of $C_V$, $\kappa_T$ and $C_s^2$. }
		\label{sqrts}
	\end{figure}

	In the event-by-event analysis, temperature can be extracted from transverse momentum ($p_T$) spectra. From our plot of $C_V$ (Fig.~\ref{sqrts}(a)) we note that the specific heat drops a bit sharply at a low value of $\sqrt{s}$ and then it almost saturates. When the centre of mass energy of heavy-ion collision is increased the colliding nuclei become more transparent and they create smaller particle number density. Hence the energy pumped into the system goes to increase the temperature and specific heat decreases with increasing centre of mass energy. We further note that the specific heat is suppressed due to repulsive interactions.\ The temperature distribution of event-by-event collisions become broadly peaked in presence of repulsive interactions. However, repulsive interaction between hadrons does not change the qualitative behaviour of $C_V$. A similar finding was observed in Ref.~\cite{Basu:2016ibk} in which authors studied the energy dependence of specific heat  of hadronic matter at freeze-out in Au+Au and Cu+Cu collisions at the  Relativistic Heavy Ion Collider (RHIC) energies within the  ambit of hadron resonance gas model. These authors found a sharp drop in $C_V$ from low collision energy till $\sqrt{s}=62.4$ GeV which then  saturates. Here one should note that the freeze-out  parameters are different in their work and as a result the numerical values of $C_V$ are different.

	In the context of HIC experiments, isothermal compressibility ($\kappa_T$) is related to the fluctuation in the particle multiplicities, temperature and volume of the system formed. In the grand canonical ensemble framework the particle multiplicity fluctuation $\omega$ is given by
	\begin{eqnarray}
	\omega=\frac{\langle N^2 \rangle- \langle N \rangle^2}{\langle N \rangle} 
	=\frac{k_B T\langle N \rangle}{V}\kappa_T
	\end{eqnarray}
	In the above equation, temperature and volume can be extracted from the mean hadron yield and $\kappa_T$ can be extracted through the event-by-event multiplicity fluctuation measurements. 
	Here we study the beam-energy dependence of $\kappa_T$ which would provide an information about its magnitude at freeze-out.

	In Fig.\ref{sqrts}(b) it can be seen that the compressibility sharply decreases till $\sqrt{s}=25$ GeV, and then it saturates. This indicates  that the matter produced in low energy collision experiments is more compressible. The effect of repulsive interactions on the magnitude of $\kappa_T$ is quite strong at higher collision energies. The magnitude of $\kappa_T$ for low collision energies is not very different from ideal HRG estimation, while at high collision energies its magnitude is higher as compared to ideal HRG estimations. However, the repulsive interaction does not change the overall behaviour of $\kappa_T$. In 
	Ref.~\cite{Khuntia:2018non,Mukherjee:2017elm} authors carried out similar studies and, qualitatively, the findings are same as ours.

	In the relativistic hydrodynamic simulation of the matter produced in HIC experiments the speed of sound plays a very crucial role.  The value of $C_s^2$ depends on the degrees of freedom and the interactions. Study of $C_s^2$ carried out in this work has vindicated this fact. We have seen how the behaviour of $C_s^2$ changes as new degrees of freedom (massive resonances and baryons) enter the system. We have also shown the effect of the inclusion of repulsive interactions. In the hydrodynamical description of matter produced in HIC,  $C_s^2$ sets the expansion time scale as  $\tau_{\text{exp}}^{-1}\sim\frac{1}{\epsilon}\frac{d\epsilon}{d\tau}=\frac{1+c_s^2}{\tau}$~\cite{Mohanty:2003va}. The thermal equilibrium is maintained only if $\tau_{\text{exp}}>\tau_{\text{coll}}$, where $\tau_{\text{coll}}$ is the collision time. 
	The $C_S^2$ rises very rapidly 
	at low $\sqrt{s}$ and then it saturates at higher collision energies. Repulsive interaction has quite a strong effect. The magnitude of $C_S^2$ is larger as compared to ideal HRG model as the effect of repulsion increases. As $C_S^2$ increases with the introduction of repulsive interaction, the expansion time scale of the system decreases. Thus it becomes more difficult to maintain the thermal equilibrium.  In Ref.~\cite{Mohanty:2003va} authors discussed the sensitivity of the hadronic spectra on  EoS obtained within the ambit of hadron resonance gas model. They found that in fixing the value of freeze-out temperature at $T\sim 120$ MeV from the transverse momentum ($p_T$) spectra, the value $C_s^2=1/5$ gives good description of the data for Landau's hydrodynamical model. This value is smaller than ideal gas EoS limit $1/3$. Thus the presence of interactions lowers the value of speed of sound and hence for such system the maintenance of thermal equilibrium is easier. However, we find in the present work that the  speed of sound is higher in magnitude as compared to ideal HRG model when repulsive interactions are taken into account. This has very non-trivial effect on thermalisation.  The repulsive interaction would not be able to maintain thermal equilibrium for a long time in the expanding plasma. It would be interesting to study the hydrodynamical evolution with EoS which includes the effect of repulsive interactions and its effect on the thermalisation.

	\section{Summary and Conclusion}
	\label{secV}
	In this work we have investigated, in the ambit of HRG mean field model, three important thermodynamic quantities which can be used to indicate the phase transition between hadronic and quark-gluon plasma matter: (i) specific heat at constant volume ($C_v$), (ii) isothermal compressibility ($\kappa_T$) and (iii) the speed of sound ($C_s^2$).
	
	We found that, in case of ideal HRG model,  specific heat increases with increasing temperature which is expected because of the abundant production of heavier hadrons. Nonetheless, HRG model does not show critical behaviour due to lack of exponential rise in the spectral density needed to show such critical behaviour.  The effect of repulsive interaction is to reduce the magnitude of $C_V$.  If the repulsive interactions are included in HRG model using mean-field approach then this model reproduces the overall behaviour of $C_V$ in good agreement with LQCD data. We further found that the effect of repulsive interactions on $\kappa_T$ is very mild, while  its role in case of $C_s^2$  is very nontrivial. With the inclusion of repulsive interactions using mean-field approach, $C_S^2$ is found to be in good agreement with the LQCD data over wide range of temperatures and baryon chemical potentials.    So, the repulsive interactions have a very strong bearing on $C_V$ and $C_s^2$.
	
	We finally discussed the implications of our results in the context of heavy-ion collision experiments. We found that both $C_V$ and $\kappa_T$ are large at low $\sqrt{s}$, then drop sharply as the collision energy increases, and finally saturates at constant value at 
	large $\sqrt{s}$. On the other hand $C_s^2$ shows opposite behaviour: its value being small at low $\sqrt{s}$, then rises very rapidly as the collision energy increases, and finally saturates to a constant value at high $\sqrt{s}$. Thus, in the context of HIC experiments, all of these quantities found to be strongly sensitive to the strength of repulsive interactions.
	
	In conclusion, the role of repulsive interactions is very important to understand the QCD phase transition since the observables, sensitive to such phenomena, show strong dependence on hadronic repulsive interactions.
	
	

	\section{Acknowledgement}
	
	G. K. is financially supported by DST-INSPIRE Faculty research grant number DST /INSPIRE/04/2017/002293. G.K. also thanks Amaresh Jaiswal for useful discussion.  S.P. is financially supported by UGC, New Delhi.	AB thanks Alexander von Humboldt (AvH) foundation and Federal 
	Ministry of Education and Research (Germany) for support 
	through Research Group Linkage programme.

\end{document}